\documentclass[twocolumn,shortnote]{jpsj2} %% for short notes
\def\degangle{\kern-.2em\r{}}

\title{
Correlation between $T_{\mbox{c}}$ and Lattice Parameters of Novel Superconducting Sodium Co Oxide Hydrate
}

\author{
Hiroya \textsc{Sakurai}$^{1}$\thanks{SAKURAI.Hiroya@nims.go.jp}, Kazunori \textsc{Takada}$^{2}$, Takayoshi \textsc{Sasaki}$^{2}$, Fujio \textsc{Izumi}$^{2}$, Ruben A. \textsc{Dilanian}$^{2}$, and Eiji \textsc{Takayama-Muromachi}$^{2}$
}

\inst{
$^{1}$Superconducting Materials Center, National Institute for Materials Science (SMC/NIMS), 1-1 Namiki, Tsukuba, Ibaraki 305-0044, Japan.\\
$^{2}$Advanced Materials Laboratory, National Institute for Materials Science (AML/NIMS), 1-1 Namiki, Tsukuba, Ibaraki 305-0044, Japan.
}

\begin{document}
\maketitle
Na$_{x}$(H$_{3}$O)$_{y'}$CoO$_{2}\cdot y''$H$_{2}$O is one of the most attractive compounds because of unconventional superconductivity induced in a triangular lattice\cite{Nature}. This compound is synthesized through a soft chemical process; parts of Na ions are deintercalated from the parent oxide of $\gamma$-Na$_{x}$CoO$_{2}$, and then the obtained sample is immersed in water. Recently, it has been elucidated that, at the latter step, H$_{3}$O$^{+}$ ions are inserted substituting the Na$^{+}$ ions between two CoO$_{2}$ layers in addition to water molecules, and the Co ions are partly reduced by water\cite{JMC}. The deintercalation of Na$^{+}$ causes the oxidation of the Co ions, and the intercalation of the H$_{2}$O molecules results in a significant expansion of $c$, enhancing two-dimensionality (2D). However, the essential stacking sequence of oxygen layers and Co layers is kept unchanged before and after the soft chemical process.

 The water content, $y\equiv y'+y''$, of Na$_{x}$(H$_{3}$O)$_{y'}$CoO$_{2}\cdot y''$H$_{2}$O can be reduced to 0.7 from 1.3 by reducing the humidity in the atmosphere\cite{JSSC}. The distance between adjacent Co layers is shortened from 9.8\AA\ in $y=1.3$ to 6.9\AA\ in $y=0.7$ and the triple layer of H$_{2}$O-Na(H$_{3}$O)-H$_{2}$O in the $y=1.3$ phase is substituted by a single layer consisting of Na(H$_{3}$O) and H$_{2}$O in the $y=0.7$ phase.  No superconductivity is observed in the $y=0.7$ phase\cite{JSSC,SakuraiPhysC}, although the structure of the $y=0.7$ phase seems to be enough two-dimensional. Thus, it is interesting to explore whether Na$_{x}$(H$_{3}$O)$_{y'}$CoO$_{2}\cdot y''$H$_{2}$O with a different stacking sequence shows superconductivity.

Very recently, we discovered a new cobalt oxide superconductor\cite{aNCO}. Its chemical formula, Na$_{x}$(H$_{3}$O)$_{y'}$CoO$_{2}\cdot y''$H$_{2}$O, is the same as that of the previous superconductor, but the stacking sequence of the CoO$_{2}$ layers is different. The new superconductor is synthesized from $\alpha$-NaCoO$_{2}$ with an O3-type structure\cite{PhysicaB}, while the previous one from $\gamma$-Na$_{0.7}$CoO$_{2}$ with a P2-type structure. As a result, the new superconductor has a P3-type structure because of gliding of the CoO$_{2}$ layers during the soft chemical process\cite{aNCO}, while the previous one has a P2-type structure.

\begin{figure}
\begin{center}
\includegraphics[width=7cm,keepaspectratio]{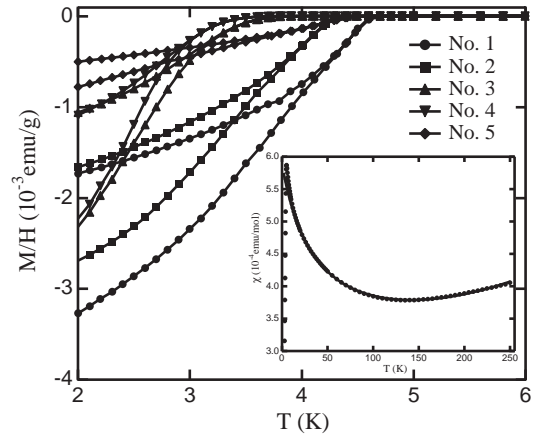}
\end{center}
\caption{
The temperature dependence of $M/H$ measured under $H=10$ Oe. The data of each sample were measured under zero-field and field cooling conditions. The inset shows $M/H$ of the No. 1 sample measured under $H=10$ kOe. The solid line shows the fitted function (see the text).
}
\label{chi}
\end{figure}

\begin{figure}
\begin{center}
\includegraphics[width=7cm,keepaspectratio]{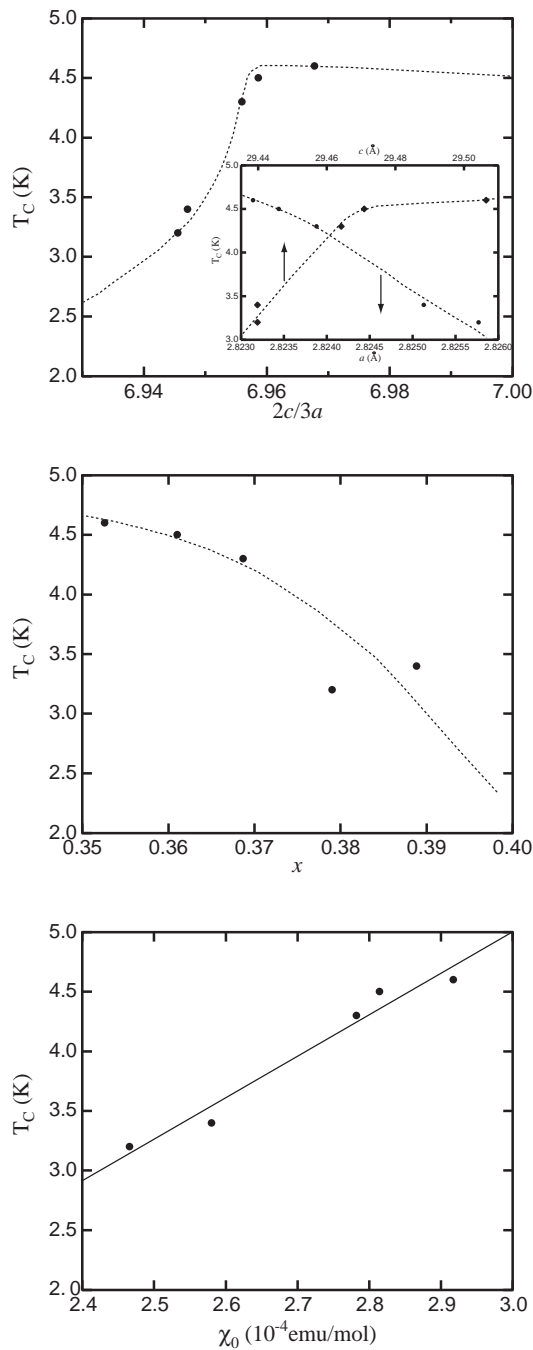}
\end{center}
\caption{
(a) The $\frac{2}{3}c/a$-dependence of $T_{\mbox{c}}$. The dotted line is visual guide\cite{Milne}. The inset shows the $a$- and $c$-dependence of $T_{\mbox{c}}$. The dotted lines are visual guides. (b) The $x$-dependence of $T_{\mbox{c}}$. The dotted line is visual guide. (c) The $\chi _{0}$-dependence of $T_{\mbox{c}}$. The solid line shows $T_{\mbox{c}}=-5.43+3.48\times 10^{4}\chi _{0}$.
}
\label{Para}
\end{figure}

In this study, we prepared several batches of samples of the new superconductor. Their chemical, structural, magnetic and superconducting properties varied rather widely in spite of essentially the same synthesis conditions. We report here some correlations between the superconducting transition temperature, $T_{\mbox{c}}$, and other properties.

The parent oxide of $\alpha$-NaCoO$_{2}$ was synthesized from a stoichiometric mixture of Na$_{2}$O$_{2}$ and Co$_{3}$O$_{4}$\cite{aNCO}. Four pellets of the mixture were fired at 600 \degC\ to obtain samples with batch numbers of 1-4. One pellet was fired at 500 \degC\ to obtain a sample with batch number 5. The average grain size of the No. 5 batch was only a few $\mu m$ and was much smaller than that in the other batches. The parent oxide underwent the soft chemical modification to become superconducting according to the method reported previously\cite{Nature}. The results of the sample characterization are shown in Table \ref{Analysis}. The ways of the characterization are the same as those reported in the ref. \citen{aNCO}.

The dependence of $M/H$ on temperature ($T$) is shown in Fig. \ref{chi}. All the samples exhibit superconductivity, but $T_{\mbox{c}}$ varies rather widely. From the zero-field cooling susceptibility data, the superconducting volume fraction was calculated at about 13\% for sample No. 1 having the highest $T_{\mbox{c}}$ of 4.6 K, and the fraction tends to decrease with decreasing $T_{\mbox{c}}$ except for sample No. 5. The transition in the No. 5 sample is sharp being comparable to that in sample No. 1 or 2, and its unreasonably small volume fraction may be caused by the small average grain size, because the system probably has a long penetration depth\cite{SakuraiPRB,Higemoto}. 

\begin{table*}[b]
\begin{center}
\begin{tabular}{||c||c|c|c|c|c|c|c|c|c|c|c|c||}\hline
No.	&	$N$	&	$C$	&	$V$	&	$x$	&	$y$	&	$a$(\AA)	&	$c$(\AA)	&	$T_{\mbox{c}}$	&	$C$	&	$\theta$	&	$A$	&	$\chi _{0}$	
\\	\hline\hline
1	&	6.52	&	47.4	&	3.48	&	0.353	&	1.40	&	2.8231	&	29.506	&	4.6	&	1.12	&	$-37.8$	&	1.22	&	2.92	
\\	\hline
2	&	6.76	&	47.0	&	3.52	&	0.369	&	1.44	&	2.8238	&	29.464	&	4.3	&	1.33	&	$-46.7$	&	1.35	&	2.78	
\\	\hline
3	&	7.16	&	47.2	&	3.51	&	0.389	&	1.39	&	2.8251	&	29.439	&	3.4	&	1.34	&	$-47.2$	&	1.40	&	2.58	
\\	\hline
4	&	6.92	&	46.8	&	3.48	&	0.379	&	1.46	&	2.8257	&	29.439	&	3.2	&	1.56	&	$-53.8$	&	1.46	&	2.47	
\\	\hline
5	&	6.69	&	47.5	&	3.47	&	0.361	&	1.38	&	2.8234	&	29.470	&	4.5	&	1.20	&	$-37.9$	&	1.30	&	2.81	
\\	\hline
\end{tabular}
\end{center}
\caption{
Parameters obtained by sample characterization and estimated from magnetic measurements. $N$ and $C$ represent mass percentages of Na and Co, respectively, while $V$ Co valence. The units of $T_{\mbox{c}}$, $C$, $\theta$, $A$, and $\chi _{0}$ are [K], [$10^{-2}$emu$\cdot$K/mol], [K],  [$10^{-9}$emu/mol$\cdot$K$^{2}$], and [$10^{-4}$emu/mol], respectively.
}
\label{Analysis}
\end{table*}

As shown in the inset of Fig. \ref{chi}, the magnetic susceptibility of sample No. 1 above $T_{\mbox{c}}$ shows the upturn below approximately 100 K as in the case of the previous superconductor with P2 structure\cite{SakuraiPRB,SakuraiPhysC}. Thus, the following equation was fit to the data between 20 K and 250 K:
$\chi = C/(T-\theta) + AT^{2}+ \chi _{0}$.
Here the first term represents the Curie-Weiss term with the Curie constant, $C$, and the Weiss temperature, $\theta$. The second term represents the $T$-dependent Pauli paramagnetism term, while the last one is the constant term. The obtained values were listed in Table \ref{Analysis}. Since these values are comparable to those obtained for the P2 phase\cite{SakuraiPRB,SakuraiPhysC}, the superconductivity of this compound is probably unconventional as in the case of the P2 phase\cite{Ishida}.

$T_{\mbox{c}}$, $x$, and the lattice parameters seem to be somewhat correlated with each other. $T_{\mbox{c}}$ decreases with decreasing lattice parameter $c$, which indicates that the 2D nature of the present system induces superconductivity. Furthermore, as shown in Fig. \ref{Para}, $T_{\mbox{c}}$ seems to depend on $\frac{2}{3}c/a$, as suggested previously for the P2 phase\cite{Milne}. Figure \ref{Para} also suggests rather clear correlation between $T_{\mbox{c}}$ and $x$; i.e. $T_{\mbox{c}}$ decreases with increasing $x$, in contrast to the report by Milne $et$ $al.$\cite{Milne} that $T_{\mbox{c}}$ is independent of $x$ in the P2 phase. It was hard to see the correlation between $T_{\mbox{c}}$ and the Co valence probably because of the insufficient accuracy of the Co valence, which needed to be determined from both the results of the redox titrations and the mass percentages of Co. However, from the magnetic parameters (in particular from $\chi _{0}$), it seems very likely that the density of state at Fermi level varies from sample to sample showing a certain correlation with $T_{\mbox{c}}$.

In summary, we synthesized the five batches of the samples of the novel P3 type superconductor, Na$_{x}$(H$_{3}$O)$_{y'}$CoO$_{2}\cdot y''$H$_{2}$O, by the soft chemical process starting from $\alpha$-NaCoO$_{2}$. The chemical and structural properties varied rather widely from batch to batch, with a result that $T_{\mbox{c}}$ varied from 4.6 K to 3.2 K. The magnetic susceptibility above $T_{\mbox{c}}$ shows upturn at low temperature as in the case of the P2 phase. The $T_{\mbox{c}}$ seems to be well correlated to the lattice parameters.

Special thanks to S. Takenouchi (NIMS). This study was partially supported by CREST of Japan Science and Technology Agency (JST) and by Grants-in-Aid for Scientific Research (B) from Japan Society for the Promotion of Science (16340111). One of the authors (H.S) is research fellow of the Japan Society for the Promotion of Science.

\end{document}